\documentclass[prb,twocolumn,showpacs,amsmath,superscriptaddress,psfig,amssymb]{revtex4-1}

\usepackage{graphicx}
\usepackage{dcolumn}
\usepackage{bm}
\usepackage{xfrac}
\usepackage{booktabs}
\usepackage{threeparttable}

\begin{document}


\title{Anisotropic superconductivity in the non-centrosymmetric BiPd}
\author{L. Jiao}
\affiliation{Center for Correlated Matter and Department of Physics, Zhejiang University, Hangzhou,
Zhejiang 310027, China}
\author{J. L. Zhang}
\affiliation{Center for Correlated Matter and Department of Physics, Zhejiang University, Hangzhou,
Zhejiang 310027, China}
\author{Y. Chen}
\affiliation{Center for Correlated Matter and Department of Physics, Zhejiang University, Hangzhou,
Zhejiang 310027, China}
\author{Z. F. Weng }
\affiliation{Center for Correlated Matter and Department of Physics, Zhejiang University, Hangzhou,
Zhejiang 310027, China}
\author{Y. M. Shao}
\affiliation{Center for Correlated Matter and Department of Physics, Zhejiang University, Hangzhou,
Zhejiang 310027, China}
\author{J. Y. Feng}
\affiliation{Center for Correlated Matter and Department of Physics, Zhejiang University, Hangzhou,
Zhejiang 310027, China}
\author{X. Lu}
\affiliation{Center for Correlated Matter and Department of Physics, Zhejiang University, Hangzhou,
Zhejiang 310027, China}
\author{B. Joshi}
\affiliation{Tata Institute of Fundamental Research, Homi Bhabha Road, Colaba, Mumbai 400005, India}
\author{A. Thamizhavel}
\affiliation{Tata Institute of Fundamental Research, Homi Bhabha Road, Colaba, Mumbai 400005, India}
\author{S. Ramakrishnan}
\affiliation{Tata Institute of Fundamental Research, Homi Bhabha Road, Colaba, Mumbai 400005, India}
\author{H. Q. Yuan}
\email{hqyuan@zju.edu.cn}
\affiliation{Center for Correlated Matter and Department of Physics, Zhejiang University, Hangzhou,
Zhejiang 310027, China}

\begin{abstract}

We report measurements of London penetration depth $\lambda(T)$ for the noncentrosymmetric superconductor BiPd by using a tunnel diode oscillator. Pronounced anisotropic behavior is observed in the low-temperature penetration depth; the in-plane penetration depth $\lambda_{ac}(T)$ follows an exponential decrease, but the interplane penetration depth $\lambda_b(T)$ shows power-law-type behavior. The superfluid density $\rho_s(T)$, converted from the penetration depth $\lambda(T)$, is best fitted by an anisotropic two-band BCS model. We argue that such a complex order parameter is attributed to the admixture of spin-singlet and spin-triplet pairing states as a result of antisymmetric spin-orbit coupling in BiPd.
\end{abstract}


\maketitle

Considerable attention has been devoted to the study of noncentrosymmetric (NCS) superconductors (SCs) and their exotic properties in recent years.\cite{Book} The absence of an inversion symmetry introduces an asymmetric potential gradient and, therefore, yields an antisymmetric spin-orbit coupling (ASOC). The ASOC may split the electron bands by lifting the spin degeneracy, allowing admixture of spin-singlet and spin-triplet pairing states within the same orbital channel.\cite{Frigeri} Furthermore, it was recently proposed that NCS SCs with a strong SOC are potential candidates for realizing topological superconductivity.\cite{Chadov}

 Currently, the effect of broken inversion symmetry on superconductivity remains a puzzle and the role of ASOC on the superconducting pairing state is still highly controversial.\cite{Chen 2013} Unconventional superconductivity was observed in the heavy fermion superconductors CePt$_3$Si,\cite{Bauer04,Yogi} CeRhSi$_3$,\cite{Kimura} and CeIrSi$_3$,\cite{Sugitani} as well as in Li$_2$Pt$_3$B.\cite{Yuan06} On the other hand, BCS-like superconductivity was claimed in a number of weakly correlated NCS SCs with heavy atoms, e.g., Re$_3$W,\cite{Zuev} Mg$_{10}$Ir$_{19}$B$_{16}$,\cite{Klimczuk} BaPtSi$_3$\cite{Bauer09} and Nb$_x$Re$_{1-x}$.\cite{Karki11} The determinant parameter remains to be revealed for the pairing states of NCS SCs. Theoretically, multiband superconductivity with anisotropic gaps or even a nodal gap structure is expected for NCS SCs, in particular when the band splitting $E_{\textrm{ASOC}}$ becomes comparable or even larger than the superconducting gap.\cite{Frigeri} To reveal the anisotropic gap structure in NCS SCs, high-quality single crystals are necessary. Unfortunately, most of the previous measurements were performed on polycrystalline samples, which restricted our studies on their complex gap symmetry.

Recently, a new noncentrosymmetric superconductor, BiPd, was successfully synthesized.\cite{Joshi} This compound undergoes a structural transition from $\beta$-BiPd (orthorhombic,Cmc2$_1$) to $\alpha$-BiPd (monoclinic, P2$_1$) at 210$^{\circ}$C and then becomes superconducting at $T_c \simeq$ 3.7K.\cite{Joshi} In comparison with many other NCS SCs, BiPd is a weakly correlated compound possessing a heavy atom Bi. Furthermore, the sample quality for BiPd is much higher than many other NCS SCs investigated to date. These unique characteristics provide us with a great opportunity to study the exotic pairing state of NCS SCs. Measurements of point contact Andreev reflection (PCAR) spectra showed evidence of multiple superconducting gaps with a zero-bias conductance peak (ZBCP) in BiPd.\cite{Mondal} Moreover, a suppressed coherence peak was observed in the recent nuclear quadrupole resonance (NQR)  experiments.\cite{ZhengGQ} These results indicate a complex gap structure in BiPd, which might be caused by the ASOC effect as a result of lacking inversion symmetry. Further experimental evidence is highly desired in order to reveal its order parameter and the underlying pairing mechanism.

In this Rapid Communication, we probe the gap symmetry of BiPd by measuring the London penetration depth down to 50mK with a tunnel diode oscillator (TDO)-based technique. Pronounced anisotropic behavior is observed for the in-plane [$\lambda_{ac}(T)$] and out-of-plane [$\lambda_{b}(T)$] penetration depth. Detailed analysis of the superfluid density $\rho_s(T)$ suggests two-gap superconductivity with anisotropy for BiPd.

Single crystals of BiPd were synthesized by a modified Bridgman method as described elsewhere.\cite{Joshi} The samples were orientated by a Laue photography method and then cut into small pieces with a typical size of 0.3--0.9 mm$^2$ $\times$ 0.2mm, with the plane being parallel or perpendicular to the $\textit{b}$ axis. Temperature dependence of the penetration depth was precisely measured by utilizing a tunnel diode oscillator mounted on a $^{3}$He cryostat or a $^3$He/$^4$He dilution refrigerator.\cite{Van Degrift} The operating frequency of this oscillator is 7 MHz with a frequency resolution as low as 0.05 Hz, which corresponds to a resolution of penetration depth of $\sim$0.1 $\textrm{\AA}$. The penetration depth change is proportional to the shift of the TDO frequency, i.e., $\Delta\lambda$(T) = $G$$\Delta f(T)$, where the \textit{G} factor is solely determined by the sample geometry.\cite{ProzorovReview}

\begin{figure}[b]\centering
  \includegraphics[width=7.6cm]{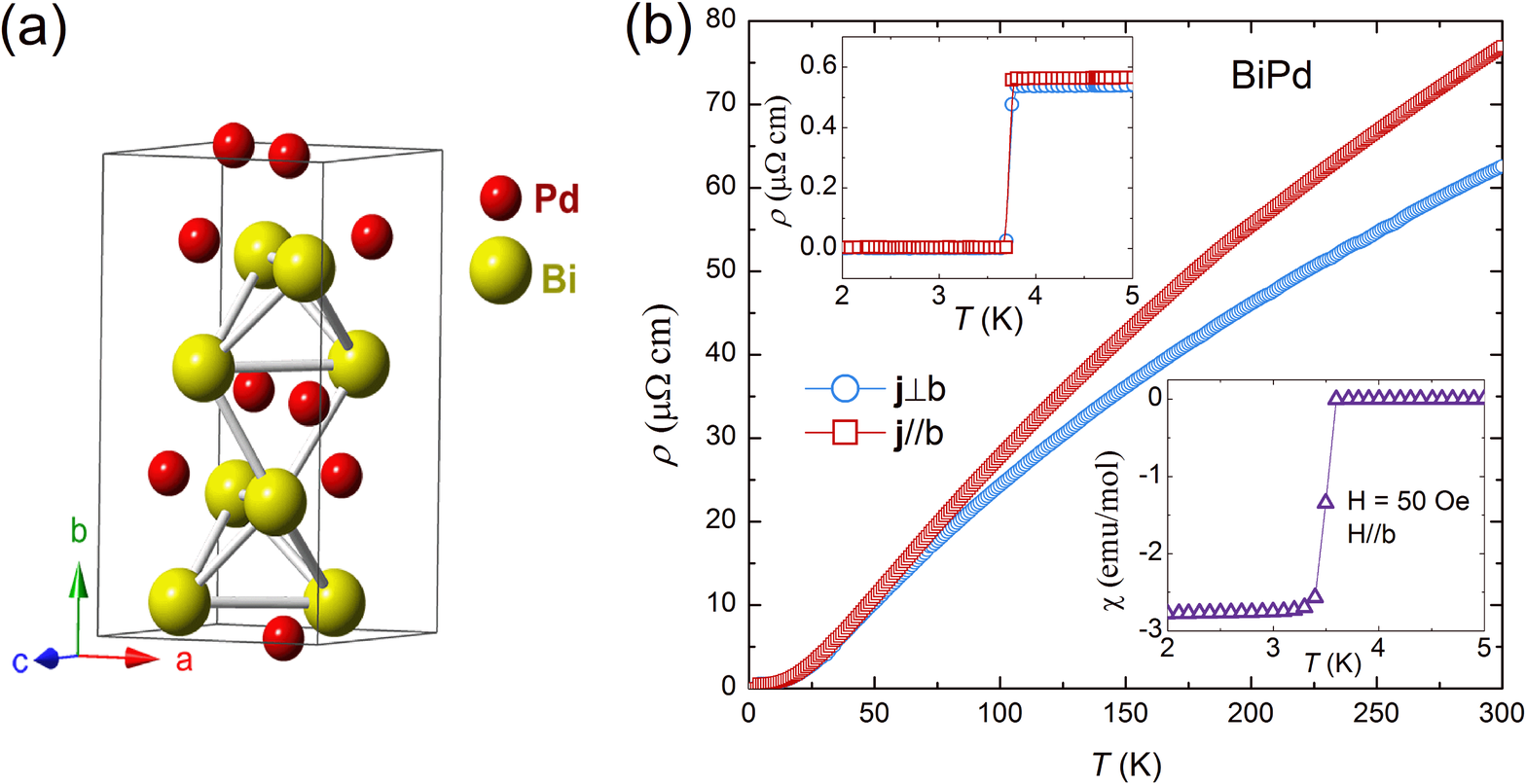}
\caption{(Color online) (a) The crystal structure of BiPd with the \textit{b} axis as its unique axis. (b) Temperature dependence of
the electrical resistivity $\rho(T)$ for BiPd with \textbf{j}$\perp$$b$ and \textbf{j}$\parallel$$b$, respectively.
The insets show $\rho(T)$ and the magnetic susceptibility $\chi(T)$ near $T_c$. The samples are from the same batch as those used for the penetration depth measurement.}\label{Fig1}
\end{figure}

As shown in Fig.~\ref{Fig1}(a), BiPd crystallizes in a monoclinic structure at low temperatures with the $b$ axis being its unique axis. The lattice constants are $a = 5.63\textrm{\AA}$, $b = 10.66\textrm{\AA}$, $c=5.68\textrm{\AA}$, $\alpha$ = $\gamma$ = 90$^{\circ}$, and $\beta$ = 101$^{\circ}$. To characterize the sample quality, we have measured the electrical resistivity $\rho(T)$ and magnetic susceptibility $\chi(T)$, which demonstrate simple metallic behavior without any magnetic order at temperatures above $T_c \simeq 3.7$ K [see Fig.~\ref{Fig1}(b)]. The sharp superconducting transition, evidenced in the electrical resistivity (top inset) and the magnetic susceptibility (bottom inset), together with a large residual resistivity ratio of RRR = 110 for \textbf{j}$\perp$$b$ and RRR = 140 for \textbf{j}$\parallel$$b$, ensures a high sample quality. The mean free path ($l \simeq$ 2422nm), estimated from the small residual resistivity [$\rho$(4 K) = 0.56$ \mu\Omega$ cm], is much longer than the coherence length ($\xi \simeq$ 30nm),\cite{Joshi} indicating that the samples are in the clean and local limit. These properties make BiPd an ideal system for the study of the mixed pairing state arising from the ASOC effect.

\begin{figure}\centering
\includegraphics[width=8.0cm]{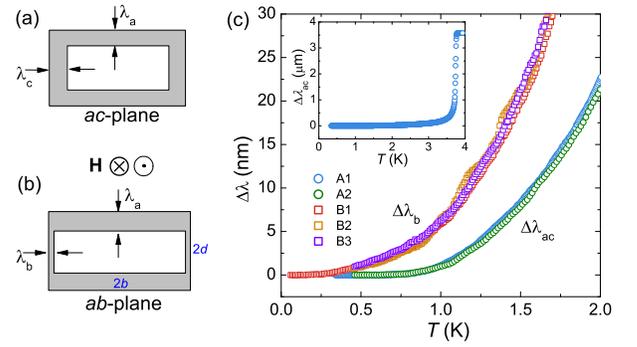}
\caption{(Color online) A schematic drawing of magnetic penetration  for (a) the isotropic \textit{ac} plane and (b) the anisotropic \textit{ab} plane, respectively. A small ac magnetic field \textbf{H} is generated perpendicular to the sample planes. The shade denotes the field-penetrating area. (c) The in-plane ($\Delta\lambda_{ac}$) and out-of-plane ($\Delta\lambda_b$) penetration depth at low temperatures for several BiPd crystals. The $G$ factors are 3.0 $\textrm{\AA}$/Hz and 5.8 $\textrm{\AA}$/Hz for samples \#A1 and \#A2, respectively. The inset shows $\Delta\lambda_{ac} (T)$ of sample \#A1 over a wide temperature range.}
\label{Fig2}
\end{figure}

In a superconductor with an anisotropic gap along the \textit{b}-axis, quantitative analysis of the London penetration depth may depend on the relative orientation of the excitation field \textbf{H} with respect to the \textit{b} axis [Fig.~\ref{Fig2}(a)]. For \textbf{H}$\parallel$$b$, the screening currents are generated in the $ac$ plane, yielding an isotropic in-plane penetration depth $\lambda_{ac}(T) = \lambda_{ac}(0)+\Delta\lambda_{ac}(T)$, where $\Delta\lambda_{ac}(T)$ = $G\Delta f^{\parallel}(T)$. For \textbf{H}$\perp$$b$, the shielding current flows along both the \textit{ac}-plane and the \textit{b} axis [Fig.~\ref{Fig2}(b)]. Thus, the London penetration depth is mixed with the in-plane and out-of-plane contributions. In this case, one needs to solve the anisotropic London equation to determine the out-of-plane penetration depth $\lambda_b(T)$. For a slab of length $2w$, width $2b$, and thickness $2d$ ($w\sim b \gg d$), $\lambda_b(T)$ can be derived by numerically solving the following equation, with $\lambda_{ac}(T)$ input from an independent measurement:\cite{ProzorovReview}

\begin{equation}
\frac{\Delta f^{\perp}(T)}{\Delta f_0^{\perp}}=1-\frac{\lambda_{ac}}{d}\textrm{tanh}(\frac{d}{\lambda_{ac}})-2\lambda_bb^2\sum_{n=0}^{\infty}\frac{\textrm{tanh}(\widetilde{b}_n/\lambda_b)}{k_n^2\widetilde{b}_n^3},
\label{eq:two}
\end{equation}
where $k_n = \pi(n+1/2)$, $\widetilde{b}_n$ = $b\sqrt{(k_n\lambda_{ac}/d)^2+1}$ and $\Delta f_0^{\perp}$ is the total frequency shift upon extracting the sample out of the coil. In this context, we calculate $\lambda_b(T)$ of various samples by taking $\lambda_{ac}(T)$ of sample \#A1 as a reference.

Figure~\ref{Fig2}(c) shows the changes of the in-plane [$\Delta\lambda_{ac}(T)$] and out-of-plane [$\Delta \lambda_b(T)$] penetration depth for BiPd. For each field orientation, several samples were measured and the data are highly reproducible. We note that the samples were cut either along or perpendicular to the $b$ axis. Within the $ac$ plane, the samples are randomly aligned and the good reproducibility of $\Delta\lambda_{ac}(T)$ for samples \#A1 and \#A2 indeed suggests an isotropic behavior of the in-plane penetration depth. However, the penetration depth shows distinctly anisotropic behavior for \textbf{H}$\parallel$$b$ and \textbf{H}$\perp$$b$. The in-plane penetration depth $\Delta\lambda_{ac}(T)$ is flattened for $T$ $<$ 1K, showing exponential-type temperature dependence below 1.75 K. On the other hand, the out-of-plane penetration depth $\Delta\lambda_b(T)$ grows much faster with temperature. The inset of Fig.~\ref{Fig2}(c) plots $\Delta\lambda_{ac}(T)$ of sample \#A1 over a broad temperature region, where the sharp drop marks a superconducting transition at $T_c = 3.7$ K, which is a value that is close to that of the electrical resistivity and magnetization.

\begin{figure}\centering
\includegraphics[width=8.7cm]{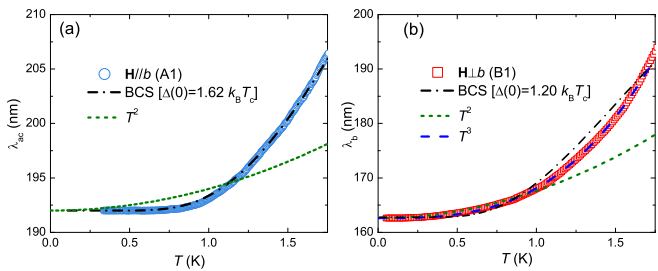}
\caption{(Color online) Temperature dependence of the (a) in-plane and (b) out-of-plane penetration depth for BiPd (symbols). The lines represent the fits of experimental data to various models.}
\label{Fig3}
\end{figure}

In order to analyze the gap symmetry, we take samples \#A1 (\textbf{H}$\parallel$$b$) and \#B1 (\textbf{H}$\perp$$b$) as examples and fit their low-temperature penetration depth with various models. Figure~\ref{Fig3} shows the temperature dependence of the penetration depth $\lambda_{ac}(T)$ and $\lambda_{b}(T)$. The penetration depth at zero temperature, $\lambda(0)$, can be estimated by $\lambda(0)$ $\approx$ 1.06*10$^{10}$/$\xi$$\gamma^{1/2}T_c$,\cite{Orlando} where $\xi$ and $\gamma$ represent the coherence length and the specific-heat Sommerfeld coefficient, respectively. By taking the values of $\xi^{\perp}$ = 32nm , $\xi^{\parallel}$ = 23nm and $\gamma$ = 4mJ/mol K$^{2}$ from the literature,\cite{Joshi} we obtain $\lambda_{b}(0)$ $\approx$ 163nm and $\lambda_{ac}(0)$ $\approx$ 192nm for BiPd, which are close to the $\mu$SR results of $\lambda(0)$ $\approx$ 230nm.\cite{Edward} In Fig. 3, we fit the penetration depth $\lambda(T)$ to the BCS model as well as the power-law behaviors. According to the isotropic BCS model in the local limit, the penetration depth can be approximated by the expression at $T \ll T_c$: $\frac{\Delta\lambda(T)}{\lambda(0)}=\sqrt{\frac{\pi\Delta(0)}{2k_BT}}\textrm{exp}(-\frac{\Delta(0)}{k_BT})$,\cite{ProzorovReview} where $\Delta$(0) is the superconducting gap amplitude at $T$ = 0. The BCS model can nicely describe $\lambda_{ac}(T)$ with $\Delta$(0) = 1.62$k_BT_c$ (0.52meV) [see Fig.~\ref{Fig3}(a)], but gives a poor fit to $\lambda_b(T)$ in the same temperature range [see Fig.~\ref{Fig3}(b)]. It is noted that, in the low-temperature limit, $\lambda_b(T)$ can be reasonably fitted by the BCS model with a small gap of 1.2$k_BT_c$ (0.38meV). Furthermore, the power-law behavior of $\lambda(T)\sim T^n$ with $n$ = 1 and 2 fails to illustrate the experimental data too, excluding nodal superconductivity for BiPd. Instead, $\lambda_b(T)$ can be reasonably fitted by $\lambda_b(T)$ $\propto T^3$ at low temperatures. Such a power-law behavior with a large exponent was previously observed in some multiband superconductors, e.g., PrPt$_4$Ge$_{12}$\cite{zjl} and Ba(Fe$_{1-x}$Co$_x$)$_2$As$_2$.\cite{BFCA} These experimental facts indicate a complex gap structure for BiPd, showing a possible scenario of multiband superconductivity with anisotropic gaps. In the following, we further elucidate this feature by analyzing the superfluid density of BiPd.

\begin{figure}\centering
\includegraphics[width=7.0cm]{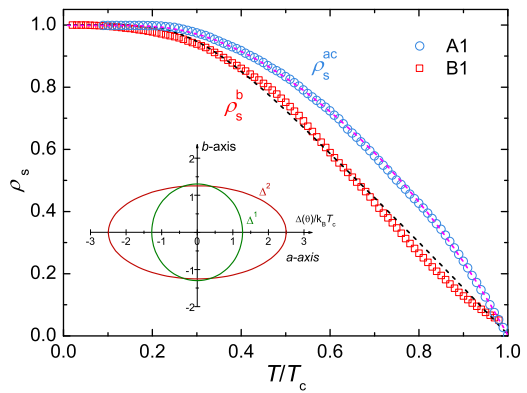}
\caption{(Color online) The two components of the superfluid density $\rho_s^{ac}$ and $\rho_s^b$ for BiPd (symbols). The dashed lines are the fits of experimental data to Eqs.~(\ref{eq:ac}) and ~(\ref{eq:b}) using a two-gap model. The inset shows the cross section of the two energy gaps, $\Delta^1$(0) and $\Delta^2$(0), in the \textit{ab} plane.}
\label{Fig4}
\end{figure}

The superfluid density $\rho_s(T)$ can be converted from the penetration depth via $\rho_s$ = $\lambda^2(0)/\lambda^2(T)$. In Fig.~\ref{Fig4}, we present the two components of the normalized superfluid density, i.e., $\rho_s^{ac}(T)$ and $\rho_s^{b}(T)$, which demonstrate anisotropic behavior as seen in the penetration depth. The behavior of superfluid density depends on the Fermi-surface topology and the gap structure. As described above, superconductivity of BiPd is fairly isotropic within the \textit{ac} plane, but becomes anisotropic along the \textit{b} axis. For simplicity, we consider a three-dimensional (3D) spheroidal gap structure while fitting the superfluid density $\rho_s(T)$,
\begin{equation}
 \Delta(T,\theta)=\frac{\Delta_{ac}(T)}{\sqrt{1-\eta \cdot z^2}},
 \label{delta}
 \end{equation}
where $z$=cos($\theta$) and $\theta$ is the polar angle with $\theta$=0 along the \textit{b} axis. The parameter $\eta$ ($-\infty\leq\eta\leq1$) is related to the eccentricity $e$, defined by $\eta$=$e^2$=1-$c^{-1}$, where $c$ is the normalized semiaxis along the $b$ axis. Temperature dependence of the superconducting gap is approximated by: $\Delta(T)=\Delta(0)\tanh \bigg(\frac{\pi k_BT_{c}}{\Delta(0)}\sqrt{(\frac{T_c}{T}-1)}\bigg)$.\cite{ProzorovReview}

Within the semiclassical approximation, the superfluid density can be calculated by:\cite{ProzorovReview}
\begin{equation}
\rho_s^{ac}=1-\frac{3}{4T}\int_0^{1}(1-z^2)\bigg[\int_0^{\infty}\textrm{cosh}^{-2}(\frac{\sqrt{\varepsilon^2+\Delta(T,\theta)^2}}{2T})d\varepsilon\bigg]dz,
\label{eq:ac}
\end{equation}
\begin{equation}
\rho_s^{b}=1-\frac{3}{2T}\int_0^{1}z^2\bigg[\int_0^{\infty}\textrm{cosh}^{-2}(\frac{\sqrt{\varepsilon^2+\Delta(T,\theta)^2}}{2T})d\varepsilon\bigg]dz.
\label{eq:b}
\end{equation}

By fitting the experimental data of $\rho_s^{ac}(T)$ and $\rho_s^b(T)$ to Eqs.~(\ref{eq:ac}) and ~(\ref{eq:b}) simultaneously, one can determine the gap parameters in Eq.~(\ref{delta}). Since the PCAR experiments have shown evidence of multiple superconducting gaps for BiPd,\cite{Mondal} here we analyze the superfluid density $\rho_s^{ac}(T)$ and $\rho_s^b(T)$ in terms of the two-band BCS model,
\begin{equation}
\rho_s^{ac,b}(T)=\omega\rho_s^{ac,b}(\Delta^1,T)+(1-\omega)\rho_s^{ac,b}(\Delta^2,T).
\label{eq:ac1}
\end{equation}
Here, $\Delta^1$ and $\Delta^2$ are defined by Eq.~(\ref{delta}). Based on the previous PCAR results,\cite{Mondal} we assume that $\Delta^1$ is isotropic, i.e., $\eta_1$ = 0. Then the free parameters in Eq.~(\ref{eq:ac1}) are $\eta_2$, $\omega$, $\Delta_{ac}^1$(0), and $\Delta_{ac}^2$(0). The best fits, as shown in Fig.~\ref{Fig4}, give parameters of $\eta_2$ = -3, $\Delta_{ac}^1$(0) = 1.3$k_BT_c$ (0.41 meV), $\Delta_{ac}^2$(0) = 2.5$k_BT_c$ (0.80 meV), and $\omega$ = 0.2. These parameters are compatible with those derived from the PCAR experiments \cite{Mondal} and the size of $\Delta_{ac}^1$(0) is also close to that derived from the fits of $\lambda_b(T)$ in the low-temperature limit. The inset of Fig.~\ref{Fig4} shows the cross section of the energy gaps $\Delta_{ac}^1$(0) and $\Delta_{ac}^2$(0) in the \textit{ab} plane. One can see that the two-band BCS model with an anisotropic gap can well describe the experimental data. It is noted that the fine gap structure relies on the Fermi-surface topology which is not yet determined for BiPd.

The anisotropic multigap superconductivity is consistent with other experiments for BiPd. For example, the reduced specific-heat jump at $T_c$ might be attributed to the effects of a multiband and/or anisotropic gap.\cite{Joshi} Furthermore, the upper critical field $\mu_0H_{c2}(T)$ shows anisotropic behavior with a pronounced concave curvature near $T_c$.\cite{Mondal} The recent NQR measurements revealed a BCS-type gap function, but with a significantly suppressed coherence peak in the spin-lattice relaxation rate;\cite{ZhengGQ} the derived gap size of $\Delta(0)$ = 1.35$k_BT_c$ (0.43 meV) is close to the small gap in our results. These experimental facts corroborate a scenario of multiband BCS superconductivity in BiPd.

\begin{table}[t]
   \caption{Pairing states and band splitting energies for several NCS SCs.}\label{tab1}
   \centering
\begin{tabular}{*6c}\hline
Compounds & $T_c$(K) & $E_{ASOC}$(meV)& $E_r$ &Paring state & Ref.\\ \hline
CePt$_3$Si &0.75 &200 & 3093 & \textit{s}+\textit{p} & \cite{Yogi,Samokhin} \\
Li$_2$Pt$_3$B & 2.6 & 200 & 892 & triplet & \cite{Yuan06,Lee} \\
BiPd &3.7 &50 & 157 & two-gap & \cite{Mondal,ZhengGQ}\\
LaNiC$_2$ & 2.75 & 42& 177 & two-gap & \cite{Chen 2013,Hase} \\
Y$_2$C$_3$ & 16 &15 & 11 & two-gap & \cite{Y2C3,Kuroiwa} \\
Li$_2$Pd$_3$B & 7.6 &30 & 46 & \textit{s}-wave & \cite{Yuan06,Lee} \\
La$_2$C$_3$ &13.2 & 30 & 26 & \textit{s}-wave & \cite{JSKim,Kuroiwa} \\
\hline
\end{tabular}
\end{table}

Several mechanisms may lead to multiband superconductivity. For example, the interband pairing might give rise to multiband BCS superconductivity if BiPd possesses multi-sheets of Fermi surface. However, the results of PCAR spectra and NQR measurements,\cite{Mondal,ZhengGQ} together with the pronounced anisotropic behavior observed in this work, seem to disfavor such a conventional scenario. On the other hand, the relatively large ASOC strength in BiPd may play an important role on its gap symmetry. Resembling the Zeeman coupling in a magnetic field, the ASOC in a NCS compound breaks the spin degeneracy of each band, giving rise to two energy bands ($E_{\vec{k}\pm}$) with different spin rotations.\cite{Frigeri} The energy of each band can be expressed as $E_{\vec{k}\pm}=\xi_{\vec{k}}\pm\alpha|\vec{g}_{\vec{k}}|$, where $\xi_{\vec{k}}$ is the band energy measured from the Fermi surface, $\alpha$ denotes the ASOC strength, and $\vec{g}_{\vec{k}}$ is a dimensionless vector. The expression of $\vec{g}_{\vec{k}}$ is determined by the detailed electronic structure. As a result of ASOC, the parity is no longer a good symmetry in NCS SCs, and the pairing state is mixed with a spin-singlet and a spin-triplet component. Accordingly, this leads to the following two gap functions:\cite{Frigeri}
\begin{equation}
\Delta_{\pm}=\psi\pm t|\vec{g}_{\vec{k}}|,
\label{eq:gap}
\end{equation}
where each gap is defined on one of the two bands formed by lifting the spin degeneracy; $\psi$ and $t$ are the spin-singlet and spin-triplet order parameter, respectively. For a sufficiently large ASOC, the interband pairing is suppressed and the spin-triplet pairing is maximized when $\vec{g}_{\vec{k}}$ is parallel to the $d$ vector of the spin-triplet order parameter. From Eq.~(\ref{eq:gap}), one can see that, even for a spherical Fermi surface, it may naturally form two superconducting gaps with a certain anisotropy in the NCS SCs. Accidental nodes may develop on $\Delta_{-}$ while the triplet component $t$ becomes dominant. Our results shown in the preceding sections are compatible with such a scenario. The two energy gaps, $\Delta^1$ and $\Delta^2$, derived from the superfluid density, may share the same origin as those of $\Delta^+$ and $\Delta^-$. In BiPd, the ASOC results in a moderate band spitting ($E_{ASOC}$ $\approx$ 50 meV; see Table~\ref{tab1}), giving rise to comparable contributions from the spin-singlet and the spin-triplet components. In this case, it is possible that two nodeless superconducting gaps with different degrees of anisotropy may develop. We shall point out that further experimental and theoretical efforts are demanded in order to elucidate its fine gap structure.

To further seek the relationship between the superconducting paring state and the ASOC strength in NCS SCs, in Table~\ref{tab1} we list the band splitting energy $E_{ASOC}$ and its ratio to $T_c$, defined as $E_r = E_{ASOC}/k_BT_c$, for several NCS SCs for which $E_{ASOC}$ values are available in literature. One can see that, except for Y$_2$C$_3$, $E_r$ serves as a good parameter to tune the mixed pairing states in NCS SCs; a large $E_r$ is usually required for a predominant spin-triplet state. In BiPd, a moderate $E_r$ value ($E_r \approx $157) was obtained,\cite{ZhengGQ} which is much smaller than that of CePt$_3$Si\cite{Samokhin} and Li$_2$Pt$_3$B,\cite{Lee} where significant contributions from a spin-triplet state have been realized, but larger than those of the BCS-like SCs, e.g., Li$_2$Pd$_3$B \cite{Lee} and  La$_2$C$_3$. \cite{JSKim} Instead, $E_r$ of BiPd is comparable to that of LaNiC$_2$, which shows strong evidence of multigap superconductivity.\cite{Chen 2013} One should emphasize that, in BiPd, the successful growth of high-quality single crystals provides us with a unique opportunity to study its anisotropy, allowing us to better understand its gap structure. Furthermore, observations of a ZBCP in the PCAR spectra \cite{Mondal} and a suppressed coherence peak by the NQR measurements \cite{ZhengGQ} also support the involvement of a spin-triplet component in the pairing state.

In summary, we have measured the temperature dependence of the London penetration depth in two orthogonal field orientations for BiPd. Anisotropic superconductivity is observed in the penetration depth and its corresponding superfluid density. For $T\ll T_c$, the in-plane penetration depth $\lambda_{ac}(T)$ shows BCS-type exponential behavior, while the out-of-plane penetration depth $\lambda_b(T)$ follows power-law-like temperature dependence. Detailed analysis of the superfluid density $\rho_s(T)$ suggests anisotropic two-band superconductivity for BiPd. As a possible scenario, these experimental results can be interpreted in terms of the mixed pairing states in NCS SCs, shedding light on superconductivity without inversion symmetry.

We are grateful to D. F. Agterberg and Q. H. Wang for helpful discussions. This work was supported by the National Basic Research Program of China (Grant No. 2011CBA00103), the National Science Foundation of China (Grant No. 10934005 and No. 11174245), the Fundamental Research Funds for the Central Universities and the Max-Planck Society under the auspices of the Max-Planck partner group of the MPI for Chemical Physics of Solids, Dresden.

\end{document}